\documentclass [a4paper,fleqn,10pt]{article}
\usepackage{graphicx}
\usepackage {subfigure,epsfig}

\usepackage {amsmath} \usepackage{amssymb} \usepackage{cite} \usepackage{amsthm}
\numberwithin{equation}{section} \numberwithin{table}{section} \mathindent=0pt
\theoremstyle{plain} \newtheorem{theorem}{Theorem}

\numberwithin{theorem}{section}

\begin{document}

\title{\textbf{Newton polygons for finding exact solutions}}
\author{Nikolai A. Kudryashov, Maria V. Demina}
\date{Department of Applied Mathematics\\
Moscow  Engineering and Physics Institute\\
(State university)\\
31 Kashirskoe Shosse,  115409\\
Moscow, Russian Federation} \maketitle

\begin{abstract}
A method for finding exact solutions of nonlinear differential
equations is presented. Our method is based on the application of
the Newton polygons corresponding to nonlinear differential
equations. It allows one to express exact solutions of the equation
studied through solutions of another equation using properties of
the basic equation itself.  The ideas of power geometry are used and
developed. Our approach has a pictorial rendition, which is is
illustrative and effective. The method can be also applied for
finding transformations between solutions of the differential
equations. To demonstrate the method application exact solutions of
several equations are found. These equations are: the Korteveg -- de
Vries -- Burgers equation, the generalized Kuramoto - Sivashinsky
equation, the fourth -- order nonlinear evolution equation, the
fifth -- order Korteveg -- de Vries equation , the modified Korteveg
-- de Vries equation of the fifth order and nonlinear evolution
equation of the sixth order for the turbulence description. Some new
exact solutions of nonlinear evolution equations are given.
\end{abstract}



\section{Introduction}

One of the most important problems of nonlinear models analysis is
the construction of their partial solutions. Nowadays this problem
is intensively studied. We know that the inverse scattering
transform \cite{Gardner01, Ablowitz01, Ablowitz03} and the Hirota
method \cite{Hirota01, Ablowitz03, Kudryashov01} are very useful in
looking for the solutions of exactly solvable nonlinear equations.
However most of nonlinear equations are nonintegrable ones. For
constructing solutions of such equations the following methods are
used: the singular manifold method \cite{Weiss01, Conte01,
Choudhary01, Kudryashov02, Kudryashov03, Kudryashov04,
Kudryashov05}, the Weierstrass function method \cite{Kudryashov05,
Kudryashov06, Yan01}, the tanh--function method \cite{Lou01,
Parkes01, Elwakil01, Fan01, Fan02, Kudryashov07}, the Jacobian
elliptic function method \cite{Liu01, Fu01, Fu02} and the
trigonometric function method \cite{Fu03, Yan02}. Most of these
methods proceed from an a priory expression for an unknown solution.
Usually this expression is a polynomial in elementary or special
functions (for example, in trigonometric, Weierstrass, Jacobian and
some other). As a result all solutions outside the given family are
lost. This disadvantage significantly decreases the strength of such
methods.

Lately it was made an attempt to generalize most of these methods
and as a result the simplest equation method appeared
\cite{Kudryashov08, Kudryashov09}. Two ideas lay in the basis of
this method. The first one was to use an equation of lesser order
with known general solution for finding exact solutions. The second
one was to take into account possible movable singularities of the
original equation. Virtually both ideas existed though not evidently
in some methods suggested earlier.

However the method introduced in \cite{Kudryashov08, Kudryashov09}
has one essential disadvantage concerning with an indeterminacy of
the simplest equation choice. This disadvantage considerably
decreases the effectiveness of the method. In this paper we present
a new approach for finding exact solutions of nonlinear differential
equations, which greatly expands the method \cite{Kudryashov08,
Kudryashov09} and which is free from disadvantage mentioned above.
The method does not postulate the simplest equation but it allows
one to find its structure using the properties of the equation
studied itself. Our method develops the ideas of power geometry
\cite{Bruno01, Bruno02, Kudryashov13, Kudryashov14}. With a help of
the power geometry we show that the search of the simplest equation
becomes illustrative and effective. Also it is important to mention
that the results obtained are sufficiently general and can be
applied not only to finding exact solutions but also to constructing
transformations for nonlinear differential equations.

The outline of this paper is as follows. The application of the
Newton polygons of nonlinear differential equations and the method
algorithm for finding exact solutions is presented in section 2. Our
method is applied to look for exact solutions of the Korteveg -- de
Vries -- Burgers equation, nonlinear evolution equation of the fifth
order and the generalized Kuramoto -- sivashinsky equations in
sections 3, 4 and 5 accordingly. Self similar solutions of the fifth
order Korteveg -- de Vries equations are considered in sections 6
and 7. Solitary and periodical waves of sixth order nonlinear
evolution equation for description of the turbulent processes are
found in sections 8 and 9.

\section{Method applied}

Let us assume that we look for exact solutions of the following
nonlinear $n$ -- order ODE
\begin{equation}
\label{e:2.1}M_n[y(z),y_z(z),y_{zz}(z),\ldots ,z]=0.
\end{equation}
In power geometry any differential equation is regarded as a sum of
ordinary and differential monomials. Every monomial can be
associated with a point on the plane according to the following
rules
\begin{equation}\label{e:2.2}
C_1\,z^{q_1}y^{q_2}$ $\longrightarrow (q_1,q_2),
\,\,\,\,\,\,\,\,\,C_2\,\frac{d^ky}{dz^k} \longrightarrow (-k,1).
\end{equation}
Here $C_1, \,C_2$ are arbitrary constants. When monomials are
multiplied their coordinates are added. The set of points
corresponding to all monomials of a differential equation form its
carrier. Having connected the points of the carrier into the convex
figure we obtain a convex polygon called the Newton polygon of
differential equation. Thus nonlinear ODE \eqref{e:2.1} can be
characterized by the Newton polygon $L_1$ on the plane. The
periphery of the polygon consists of vertexes and edges. By
$\{\Gamma_j^{(0)}\}$ let us denote the vertexes and by
$\{\Gamma_j^{(1)}\}$ the edges. Most of edges and vertexes of the
Newton polygon define power or non -- power asymptotics and power
expansions for the solutions of the equation \cite{Bruno01, Bruno02,
Kudryashov13, Kudryashov14}.

Now let us assume that a solution $y(z)$ of the studied equation can
be expressed through solutions $Y(z)$ of another equation. The
latter equation is called the simplest equation. Consequently we
have a relation between $y(z)$ and $Y(z)$
\begin{equation}\label{e:2.3}
y(z)=F(Y(z),Y_z(z),\ldots ,z).
\end{equation}
The main problem is to find the simplest equation. Substitution
\eqref{e:2.3} into basic equation \eqref{e:2.1} yields a transformed
differential equation, which is in its turn characterized by the
polygon $L_2$. By $L_3$ denote the Newton polygon corresponding to
the simplest equation. Analyzing the Newton polygon $L_2$ we should
construct the polygon  $L_3$. Let the edge $\Gamma_1^{(1)}$ belong
to $L_2$ and the edge $\tilde{\Gamma}_1^{(1)}$ belong to $L_3$. It
can be proved that if $\Gamma_1^{(1)}$ and $\tilde{\Gamma}_1^{(1)}$
have equal external normal vectors, then corresponding asymptotics
coincide accurate to the numerical parameter. Under the term
external vector we mean the vector that goes out of the Newton
polygon. Consequently, suitable Newton polygon $L_3$ has all or
certain part of edges parallel to those of $L_2$. Besides that when
the Newton polygon $L_3$ moves along the plane his apexes should
cover the support of the transformed differential equation. From
this fact we see in particular that suitable $L_3$ is of equal or
lesser area than $L_2$. Let us suppose that we have found such
polygon $L_3$. Then we can write out the simplest equation
\begin{equation}\label{e:2.4}
E_m(Y(z),Y_z(z),\ldots ,z)=0.
\end{equation}
It is important to mention that the choice of the simplest equation
is not unique. If the following correlation
\begin{equation}\label{e:2.5}
M_n(F(Y,Y_z,\ldots ,z))=\hat{R}\,E_m(Y,Y_z,\ldots ,z)
\end{equation}
(where $\hat{R}$ is a differential operator) is true then it means
that for any solution $Y(z)$ of the simplest equation \eqref{e:2.4}
there exists a solution of \eqref{e:2.1}. Generally speaking, any
nonlinear solvable differential equation can be the simplest
equation. The only requirement is the following: the order of the
simplest equation should be lesser than the order of the transformed
differential equation. But the most important simplest equations are
those that have solutions without movable critical points. If the
general solution (or a partial solution) of the simplest equation
can be found, then we get explicit representation for a solution of
the equation studied. Otherwise we have only the relation
\eqref{e:2.3} between solutions of \eqref{e:2.1} and \eqref{e:2.4}.

The most useful examples of the simplest equations are the
following: the Riccati equation
\begin{equation}\label{e:2.6}
Y_z+Y^2- a(z) \,Y-b(z)=0,
\end{equation}
the equation for the Jacobi elliptic functions
\begin{equation}\label{e:2.7}
R_z^2=-4\,R^4+a\, R^3+b\, R^2+c\, R +d,
\end{equation}
and the equation for the Weiershtrass elliptic functions
\begin{equation}\label{e:2.8}
R_z^2=-2\,R^3+a\, R^2+b\, R+c.
\end{equation}

Solutions of equations \eqref{e:2.6}, \eqref{e:2.7}, and
\eqref{e:2.8} do not have movable critical points.

Now we are able to state our method. It is composed of six steps.

\textit{The first step.} Construction of the Newton polygon $L_1$,
which corresponds to the equation studied.

\textit{The second step.} Determination of the movable pole order
for solutions of the equation studied and transformation of this
equation using the expression \eqref{e:2.3}.

\textit{The third step.} Construction of the Newton polygon $L_2$
corresponding to the transformed equation.

\textit{The fourth step.}  Construction of the Newton polygon $L_3$,
which will characterize the simplest equation. This polygon should
posses properties discussed above.

\textit{The fifth step.} Selection of the simplest equation with
unknown parameters that generates the polygon $L_3$.

\textit{The sixth step.} Determination of the undefined
coefficients, which are present in the transformation \eqref{e:2.3}
and in the simplest equation.

\textit{Remark 1.} Very often suitable simplest equations can be
found without making transformation \eqref{e:2.3}. In this case
\eqref{e:2.3} is an identity substitution and we set $y(z)\equiv
Y(z)$, $L_2\equiv L_1$.

\textit{Remark 2.} In some cases the transformation can be included
into the simplest equation. Then again \eqref{e:2.3} is an identity
substitution.

\textit{Remark 3.} The most powerful transformations, i.e. the
transformations that generate new classes of exact solutions, are
those that change the pole order of $y(z)$.

\section{Exact solutions of the Korteveg -- de Vries -- Burgers equation}

To demonstrate our method application let us find exact solutions of
the Korteveg -- de Vries -- Burgers equation
\begin{equation}
\label{3.1} u_{t}+u\,u_x+\beta u_{xxx}-\nu u_{xx}=0.
\end{equation}
It has the travelling wave reduction
\begin{equation}
\label{3.2} u(x,t)=w(z), \,\,\,\,\,\,\,z=x-C_{0}\,t,
\end{equation}
where $w(z)$ satisfies the equation
\begin{equation}
\label{3.3} E[w]= \beta w_{zz}-\nu w_z +\frac 12\,w^2 - C_0 w
+C_1=0.
\end{equation}
Here $C_0$,\, $C_1$, \,$\alpha$, and $\beta$ are constants. In the
case,
\begin{equation}
\label{3.4} C_0= -\frac{6\nu^2}{25\beta}, \,\,\,\,\,\,C_1=0
\end{equation}
equation \eqref{3.3} is integrable and was solved by Painlev\'{e}
\cite{Kudryashov01, Polyanin01}. Its general solution is expressed
via the Jacobi elliptic function. However equation \eqref{3.3} also
describes a solitary wave. The solution in the form of this solitary
wave was first obtained in \cite{Kudryashov02} and later it was
rediscovered a lot of times.

Let us formulate the following theorem.
\begin{theorem}
\label{T:5.1.} Let $Y(z)$ be a solution of the equation

\begin{equation}
\label{e:3.7a}Y_{z}\,+\,Y^2-\frac{\nu^2}{10\,\beta}=0.
\end{equation}
Then

\begin{equation}
\label{e:3.7b}y(z)=C_{0}+\frac{3\,\nu^2}{25\,\beta}-\,\frac{12\,\nu}{5}\,Y(z)-12\,\beta\,Y(z)^2
\end{equation}
is a solution of the equation \eqref{3.3} in the case,
\begin{equation}
\label{e:3.7c}C_{{1}}\,=\,\frac
{C_0}{2}-\frac{18\,\nu^2}{625\,\beta}
\end{equation}

\end{theorem}

\begin{proof} At the
first step we should find the Newton polygon $L_1$ corresponding to
equation \eqref{3.3}. For monomials of this equation we have points:
$M_1=(-2,1),\,\,\, M_2=(-1,1),\,\,\,M_3=(0,2),\,\,\,
M_4=(0,1),\,\,\, M_5=(0,0)$. The support of equation \eqref{3.3} is
defined by four points: $Q_1=M_1,\,\,\, Q_2=M_3,\,\,\, Q_3=M_5 $ and
$Q_4=M_5$. Their convex hull is the triangle $L_1$ (Fig. 1). This
triangle contains three vertexes $\Gamma_j^{(0)}=Q_j\,(j=1,2,3)$ and
three edges
$\Gamma_1^{(1)}=[Q_1,Q_2],\,\Gamma_2^{(1)}=[Q_2,Q_3],\,\Gamma_3^{(1)}=[Q_1,Q_3]$.

\begin{figure}[h]
 \centerline{\epsfig{file=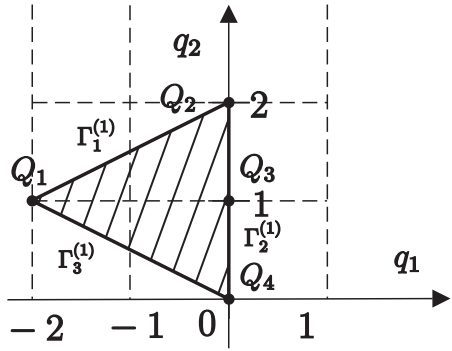,width=60mm}}
 \caption{Polygon corresponding to the differential equation \eqref{3.3}.}\label{fig1:z_post}
\end{figure}

Solutions of equation \eqref{3.3} have the second -- order
singularity. Let us make the following transformation
\begin{equation}
\label{3.5}w(z)= A_{0}+A_{1}\,Y(z)+A_2\,Y(z)^2,
\end{equation}
where $Y(z)$ satisfies the first -- order differential equation and
has the first -- order singularity.

Substituting \eqref{3.5} into the equation \eqref{3.3}, we obtain
\begin{equation}
\begin{gathered}
\label{3.6}M_{2}[w[Y]]=  \left( \beta\,A_{{1}}+2\,\beta\,A_{{2}} Y
\right) Y_{{{\it zz}}}- \left(\nu\,A_{{1}}+2\,\nu\,A_{{2
}}Y \right) Y_{{z}}+\\
\\
+ \, 2\,\beta\,A_{{2}} \, {Y_{{z}}}^{2}+
 \left(A_{{0}}A_{{2}} -C_{{0}}A_{{2}}
 +\frac{1}{2}\,{A_{{1}}}^{2} \right) {Y}^
{2}+A_{{1}}A_{{2}}{Y}^{3}+\\
\\
+\frac{1}{2}\,{A_{{2}}}^{2}{Y}^{4}+ \left(
A_{{0}}A_{{1}}-C_{{0}}A_{{1}} \right) Y+\frac{1}{2}\,{A_{{0}}}^{2
}+C_{{1}}-C_{{0}}A_{{0}}=0.
\end{gathered}
\end{equation}

The following points correspond to the monomials of this equation:
$M_1=(-2,1)$, $M_2=(-2,2)$, $M_3=(-2,2)$, $M_4=(-1,1)$,
$M_5=(-1,2)$, $M_6=(0,0)$, $M_7=(0,2)$, $M_8=(0,4)$, $M_9=(0,1)$,
$M_{10}=(0,2)$, $M_{11}=(0,3)$, $M_{12}=(0,0)$, $M_{13}=(0,1)$,
$M_{14}=(0,2)$, $M_{15}=(0,0)$. The support of equation \eqref{3.6}
is determined by seven points: $Q_1=M_1,\,\,\, Q_2=M_2=M_3,\,\,\,
Q_3=M_8,\,\,\, Q_4=M_{11},\,\,\, Q_5=M_7=M_{14}$ and
$Q_6=M_9=M_{13}$. Their convex hull is the quadrangle (Fig. 2). This
quadrangle has four vertexes $\Gamma_j^{(0)}=Q_j\,(j=1,2,3,4)$ and
four edges
$\Gamma_1^{(1)}=[Q_1,Q_2],\,\Gamma_2^{(1)}=[Q_2,Q_3],\,\Gamma_3^{(1)}=[Q_3,Q_7],\,\Gamma_4^{(1)}=[Q_1,Q_7]$.
So we have constructed the Newton polygon $L_2$.

Following our method we should find the Newton polygon $L_3$ with
some part of edges parallel to those of $L_2$. Besides that when the
polygon $L_3$ moves along the plane, his vertexes should cover the
support of equation \eqref{3.6}.
\begin{figure}[h]
 \centerline{\epsfig{file=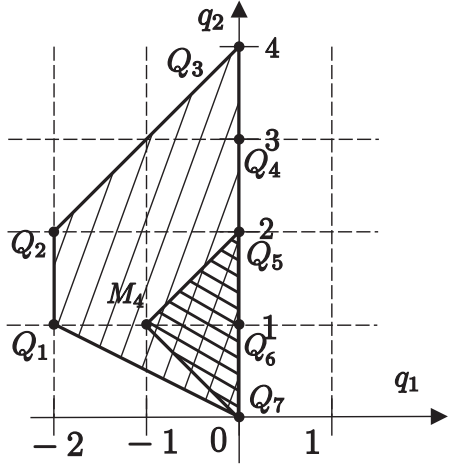,width=60mm}}
 \caption{Polygons of the differential equations \eqref{3.6} and \eqref{3.7}.}\label{fig2:z_post}
\end{figure}

We see that the triangle with vertexes $M_4\,\,\,\, Q_5$ and $Q_7$
satisfy these requirements. Thus the simplest equation is the
Riccati equation with constant coefficients
\begin{equation}
\label{3.7} E_{1}[Y]\,=\,Y_z+Y^2-b=0.
\end{equation}
Substituting
\begin{equation}
\label{e:3.8} Y_z(z)\,=\,E_1[Y]-Y^2+b
\end{equation}
into equation \eqref{3.6} and equating coefficients at powers of
$Y(z)$ to zero, yields algebraic equations for parameters $A_2$,
$A_1$, $A_0$, $b$ and $C_1$. Solving these equations, we get

\begin{equation}\begin{gathered}
\label{3.9}
A_2=-12\,\beta,\,\,\,\,A_1=-\frac{12\nu}{5},\,\,\,\,A_0=C_0+\frac{3\nu^2}{25\beta},\,\,\,\,\,b=\frac{\nu^2}{100\beta^2},\\
\\C_1=\frac12\,C_0^{2}-\frac{18\nu^4}{625\beta^2}
\end{gathered}\end{equation}
and the relation
\begin{equation}
\label{e:3.9a}M_{2}[w(Y)]\,=\hat{R}\,E_{1}[Y],
\end{equation}
where $\hat{R}$ is a differential operator. This completes the
proof.
\end{proof}

From equality \eqref{e:3.9a} we see that if $Y(z)$ is a solution of
equation \eqref{e:3.8}, then $w(z)$ in formula \eqref{3.5} is a
solution of equation \eqref{3.3}.

Hence we have found the solitary wave in the form of a kink
\cite{Kudryashov02}
\begin{equation}\begin{gathered}
\label{3.10}
w(z)=C_0+\frac{6\,\nu^2}{25\beta}-\frac{3\,\nu^2}{25\beta}\,\left(1+\tanh\left\{\pm\frac{\nu\,(x-\,C_0\,t+\varphi_0)}{10\,\beta}\right\}\right)^2.
\end{gathered}\end{equation}
Here $\varphi_0$ is an arbitrary constant.

\section{Exact solutions of the nonlinear fourth -- order evolution equation}

Let us look for exact solutions of the following fourth -- order
evolution equation \cite{Weiss02}
\begin{equation}
\label{4.1} u_{t}-2\,
u_x\,u_{xx}-u^2\,u_{xx}-2\,u\,u_x^{2}+u_{xxxx}=0.
\end{equation}
Using the travelling wave reduction \eqref{3.2} and integrating with
respect to $z$, we get
\begin{equation}
\label{4.2} w_{zzz}- w_z^{2}- w^2\,w_z-C_0\,w+C_1=0.
\end{equation}

Later let us prove the following theorem.
\begin{theorem}
\label{T:4.1.} Let $Y(z)$ be a solution of the equation

\begin{equation}
\label{4.6}Y_{zz}\,- \,Y\,Y_z-\,C_0\,=0.
\end{equation}
Then $w(z)=Y(z)$ is a solution of the equation \eqref{4.2}, provided
that $C_1=0$.

\end{theorem}

\begin{proof}Let us find the Newton polygon that corresponds to the equation
\eqref{4.2}. The following points are assigned to the monomials of
this equation: $M_1=(-3,1),\,\,\, M_2=(-2,2),\,\,\,
M_3=(-1,3),\,\,\, M_4=(0,1),\,\,\, M_5=(0,0)$. The carrier of the
equation contains five points $Q_1=M_1,\,\,\, Q_2=M_2,\,\,\, Q_3=M_3
,\,\,\, Q_4=M_4$ è $Q_5=M_5$. Their convex hull is the quadrangle
$L_1$ with four vertexes $\Gamma_j^{(0)}=Q_j\,(j=1,2,3,4)$ and four
edges
$\Gamma_1^{(1)}=[Q_1,Q_3],\,\Gamma_2^{(1)}=[Q_3,Q_4],\,\Gamma_3^{(1)}=[Q_4,Q_5],\,\Gamma_4^{(1)}=[Q_1,Q_5]$
(see Fig. 3). Solution of equation \eqref{4.2} have the first order
pole. Suitable polygon $L_3$ we find without making the
transformation. It is the triangle in Figure 3. Thus we set
\begin{equation}
\label{4.3}w(z) \equiv Y(z),
\end{equation}
where $Y(z)$ is a solution of the second order equation
\begin{equation}
\label{4.2a} E_2[Y]\,=\,Y_{zz}\,- \,a\,Y\,Y_z \,-\, b=0.
\end{equation}

\begin{figure}[h]
 \centerline{\epsfig{file=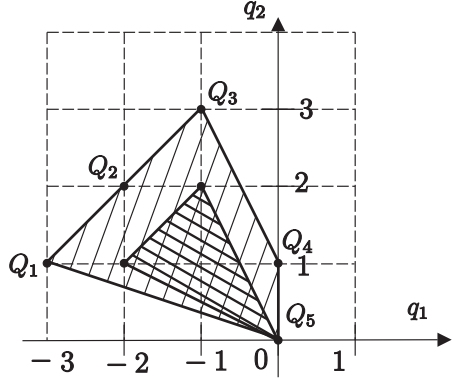,width=60mm}}
 \caption{Polygons of the differential equations \eqref{4.2} and \eqref{4.2a}.}\label{fig3:z_post}
\end{figure}

  Substituting \eqref{4.3} and
\begin{equation}
\label{4.8} Y_{zz}\,=\,E_2[Y]\,+\,a\,Y\,Y_z\,+\,b\,
\end{equation}
into equation \eqref{4.2} and equating coefficients at powers of
$Y(z)$ to zero yields algebraic equations for parameters $a$, $b$
and $C_1$. Hence we get
\begin{equation}
\label{4.8a} a=1,\,\,\,\,b=C_0,\,\,\,\,C_1=0.
\end{equation}
We also obtain the relation
\begin{equation}
\label{4.9}M_3[w(Y)]\,=\hat{R}\,E_2[Y],
\end{equation}
where $\hat{R}$ is a differential operator. This completes the
proof.
\end{proof}

Integrating equation \eqref{4.6} with respect to $z$ yields the
Riccati equation
\begin{equation}
\label{4.9b}Y_z-\,Y^2-\frac{C_0}{2}\,z+2\,C_3=0,
\end{equation}
where $C_3$ is a constant of integration. Setting
\begin{equation}
\label{4.10}Y(z)=-\frac{2\,\Psi_z}{\Psi}
\end{equation}
in \eqref{4.9b} we get
\begin{equation}
\label{4.11}\Psi_{zz}+\left(\frac{C_0}{2}\,z\,-C_3\right)\Psi=0.
\end{equation}
This equation is equivalent to the Airy equation. Its general
solution is
\begin{equation}
\label{4.12}\Psi(z)\,=\,C_1\,Ai\{-\,{2}^{-1/3}\,C_0^{1/3}\,z+C_4\}+C_2\,Bi\{-\,{2}^{-1/3}\,C_0^{1/3}\,z+C_4\},
\end{equation}
where $Ai(\zeta)$ and $Bi(\zeta)$ are the Airy functions and $C_1$,
$C_2$ and $C_4$ are arbitrary constants.

\section{Exact solutions of the generalized Kuramoto -- Sivashinsky equation}

Let us look for exact solutions of the generalized Kuramoto --
Sivashinsky equation, which can be written as

\begin{equation}\begin{gathered}
\label{e:5.0}u_t\,+\,\alpha\,u^{m}\,u_x\,+\,\delta\,u_{xx}+\,\beta\,u_{xxx}+\,\gamma\,u_{xxxx}\,=\,0.
\end{gathered}\end{equation}

Equation \eqref{e:5.0} at $m=1$ is the famous Kuramoto --
Sivashinsky equation \cite{Kuramoto01, Sivashinsky01}, which
describes turbulent processes. Its exact solutions at $\beta=0$ were
first found  in \cite{Kuramoto01}. Its solitary waves at
$\beta\neq0$ were obtained in \cite{Kudryashov02} and periodical
solutions of this equation also at $\beta\neq0$ were first presented
in \cite{Kudryashov05}. Recently it was shown \cite{Hone01,
Eremenko01} that this equation did not have other solutions except
found before.

Using the variables

\begin{equation}\begin{gathered}
\label{e:5.0a}
x'\,=\,x\,\sqrt{\frac{\delta}{\gamma}},\,\,\,\,\,\,t'=t\,\frac{\delta^2}{\gamma},\,\,\,\,\,\,u'=u,\,\,\,\,\,\,
\sigma\,=\,\frac{\beta}{\sqrt{\gamma\,\delta}}
,\,\,\,\,\,\,\alpha'\,=\,\frac{\alpha\,\sqrt{\gamma\,\delta}}{\delta\,},
\end{gathered}\end{equation}
we get the equation in the form (the primes are omitted)

\begin{equation}\begin{gathered}
\label{e:5.1}u_t\,+\,\alpha\,u^{m}\,u_x\,+\,u_{xx}+\,\sigma\,u_{xxx}+\,u_{xxxx}\,=\,0.
\end{gathered}\end{equation}

Using the travelling wave reduction
\begin{equation}\label{e:5.2}
u(x,t)=y(z),\,\,\,\,\,\,z=x-C_0\,t
\end{equation}
and integrating with respect to $z$, we get the equation
\begin{equation}\label{e:5.5}
M_3[y]\,=\,y_{zzz}\,+\sigma\,y_{zz}\,+\,y_{z}\,-\,C_0\,y\,+\,\frac{\alpha}{m+1}\,\,y^{m+1}\,=0.
\end{equation}
Here a constant of integration is equated to zero.

Let us present our result in the following theorem.

\begin{theorem}
\label{T:3.1.} Let $y(z)$ be a solution of the equation

\begin{equation}
\label{e:5.7}y_{z}\,=\left(-\frac{9\,\alpha}{2\,{m}^{3}+11\,{m}^{2}+18\,m+9}\right)^{\frac
13}\,y^{\frac{m+3}{3}}\,\mp\,{\frac {3}{\sqrt
{2\,{m}^{2}+18\,m+27}}}\,\,y.
\end{equation}
Then $y(z)$ is also a solution of the equation
\eqref{e:5.5},provided that $m\,\neq\,0$, $m\,\neq\,-1$,
$m\,\neq\,-\frac {3}{2}$, $m\,\neq\,-3$,
$m\,\neq\,\frac{-9\,\pm\,3\,\sqrt{3}}{2}$, and
\begin{equation}
\label{e:5.8}C_{{0}}=\mp\,{\frac {3\,(2\,{m}^{2}+9\,m+9)}{ \left(
2\,{m}^{2}+18\,m+27 \right) ^{3/2}}},\,\,\,\,\,\sigma=\pm\,{\frac
{3\,(3+m)}{\sqrt {2\,{m}^{2}+18\,m+27}}}.
\end{equation}

\end{theorem}

\begin{proof}
Monomials of equation \eqref{e:5.5} are determined by the following
points: $M_1=(-3,1)$, $M_2=(-2,1)$, $M_3=(-1,1)$, $M_4=(0,1)$ and
$M_5=(0,m+1)$. The Newton polygon $L_1$ that corresponds to equation
\eqref{e:5.5} is the major triangle in Figure 4 ((a): $m>0$ and (b):
$m<0$). Suitable polygon $L_3$ can be found without making the
transformation. It is the smaller triangle in Figure 4. This
triangle containes the points: $Q_1=(-1,1)$, $Q_2=M_4=(0,1)$, and
$Q_3=(0, 1+\frac{m}{3})$.

\begin{figure}[h]
 \centerline{
 \subfigure[]{\epsfig{file=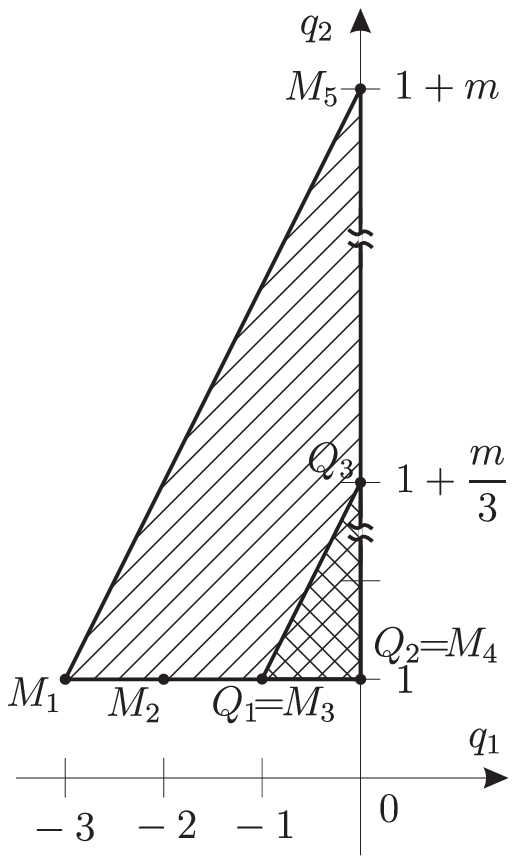,width=40mm}\label{fig4a:z_post_a}}
 \subfigure[]{\epsfig{file=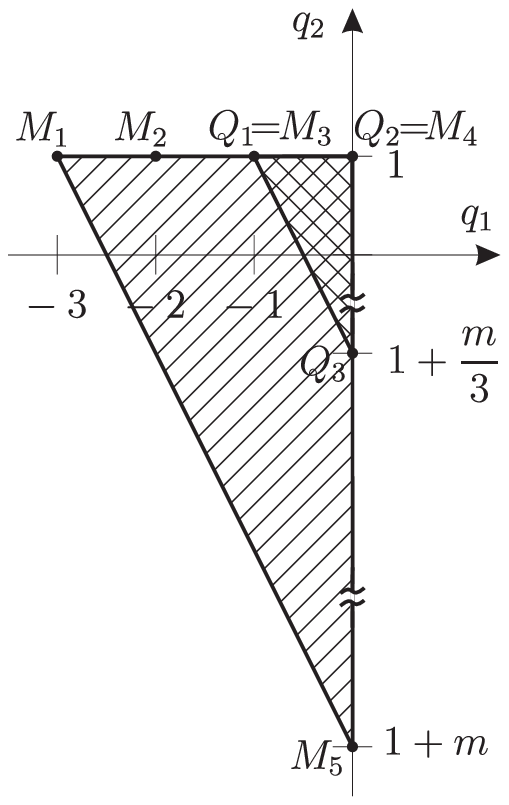,width=40mm}\label{fig4b:z_post_b}}}
 \caption{Polygons of the differential equations \eqref{e:5.5} and \eqref{e:5.6} at $m>0$ (a) and $m<0$ (b).}\label{fig4:z_post}
\end{figure}

Consequently the simplest equation corresponding to $L_3$ can be
written as
\begin{equation}
\label{e:5.6}E_1[y]\,=\,y_{z} -A\,y^{\frac{m+3}{3}}-B\,y=0,
\end{equation}
where $A$ and $B$ are parameters to be found.

The general solution of \eqref{e:5.6} takes the form
\begin{equation}
\label{e:5.6a}y(z) =\,\left(C_1\,\exp{\left\{-\frac{B\,m\,z}
{3}\right\}-\,\frac{A}{B}}\right)^{-\frac{3}{m}}.
\end{equation}

 Substituting
\begin{equation}
\label{e:5.9} y_z\,=\,E_1[y]+\,A\,y^{\frac{m+3}{3}}\,+\,B\,y
\end{equation}
into equation \eqref{e:5.5} and equating coefficients at powers of
$y(z)$ to zero yields algebraic equations for parameters $A$, $B$,
$C_0$ and $\sigma$ in the form

\begin{equation}
\label{e:5.10} {A}^{3} \left( 2\,m+3 \right)  \left(m+ 3 \right)
\left( m+1 \right) +9\,\alpha=0,
\end{equation}

\begin{equation}
\label{e:5.11} \left( m+1 \right)  \left(
{B}^{3}+\sigma\,{B}^{2}+B-{\it C_0}
 \right)\,=\,0,
\end{equation}

\begin{equation}
\label{e:5.12}{A}^{2} \left( 3+m \right)  \left( m+1 \right) \left(
B\,m+3 \,B+\sigma \right) =0,
\end{equation}

\begin{equation}
\label{e:5.13}\left( m+1 \right)  \left(
2\,{\sigma}^{2}{m}^{2}+18\,{\sigma}^{2}\,m\,+\,27\,{\sigma}^{2}\,-\,9\,{m}^{2}\,-\,54\,m\,-\,81
\right)=0.
\end{equation}
Equations \eqref{e:5.10}, \eqref{e:5.11}, \eqref{e:5.12}, and
\eqref{e:5.13} can be solved and we obtain
\begin{equation}\begin{gathered}
\label{e:5.14}A\,=(-9\,\alpha)^{1/3}\,
(\,{m}^{3}+11\,{m}^{2}+18\,m+9)^{-1/3}\\
\end{gathered}\end{equation}

(In final expressions we will take into consideration $A_1$ only.)

\begin{equation}
\label{e:5.16}C_0\,=\,B\, \left( {B}^{2}+\sigma\,B+1 \right)
,\,\,\,\,\,\,\,B\,=-{\frac {\sigma}{3+m}},
\end{equation}

\begin{equation}
\label{e:5.17}\sigma=\pm\,{\frac {3\,(3+m)}{\sqrt
{2\,{m}^{2}+18\,m+27}}}.
\end{equation}

So we have found parameters of the equation \eqref{e:5.6a},
conditions \eqref{e:5.8} and the relation
\begin{equation}
\label{e:5.18}M_3[y]\,=\hat{R}\,E_1[y],
\end{equation}
where $\hat{R}$ is a differential operator. This completes the
proof.
\end{proof}

The general solution of equation \eqref{e:5.8} can be presented in
the form

\begin{equation}
\label{e:5.19}y(z)\,=\,\left(C+C_1\,\exp{\left\{\pm{\frac{\,m\,z}{\sqrt
{2\,{m}^{2}+18\,m+27}}}\right\}}\right)^{-\frac{3}{m}},
\end{equation}
where $C_1$ is an arbitrary constant and $C$ is determined by
expression

\begin{equation}\label{e:5.20}
C=\,\pm\,{\frac {\sqrt [3]{-9\,\alpha}\,\sqrt
{2\,{m}^{2}+18\,m+27}}{3\sqrt [3]{ \left( m+3 \right)
 \left( 2\,{m}^{2}+5\,m+3 \right)} }}.
\end{equation}

We would like to note that the value of solitary wave velocity
\eqref{e:5.19} tends to $C_0\,=\,\mp\,1/\sqrt{27}$ as
$m\rightarrow\,0$, but at the same time $C_0\rightarrow\,0$ as
$m\rightarrow\infty$.

Assuming $m=1$ in \eqref{e:5.19} we obtain the known solitary wave

\begin{equation}\label{e:5.21}y(z)=\left(\frac{\sqrt
{47}\,\sqrt
[3]{-225\,\alpha}}{30}+C_1\exp{\left\{\pm{\frac{\,x\,-\,C_0\,t}{\sqrt
{47}}}\right\}}\right)^{-{3}},\,\,\,C_0=\mp\frac{60}{47\sqrt{47}}.
\end{equation}
and the value of the parameter $\sigma$: $\sigma\,=\,\pm
12/\sqrt{47}$. This solution of the Kuramoto -- Sivashinsky equation
is the kink \cite{Kudryashov02, Kudryashov03, Kudryashov04,
Kudryashov05, Kudryashov06}.

\section{Self similar solutions of the fifth -- order Korteveg -- de Vries equation}

Let us find exact solutions of the fifth -- order Korteveg -- de
Vries equation using our approach. This equation can be written as

\begin{equation}\begin{gathered}
\label{7.1}u_t +u_{xxxxx}- 10\,u\,u_{xxx}- 20\,u_x u_{xx}
 + 30\,u^2\,u_x  = 0.
\end{gathered}\end{equation}
It has the self similar solution
\begin{equation}\begin{gathered}
\label{7.2} u(x,t)=\frac{1}{(5t)^{2/5}} \,y(z),\,\,\,\,\,\,
z=\frac{x}{(5t)^{1/5}},
\end{gathered}\end{equation}
where $y(z)$ satisfies the nonlinear ODE of the form

\begin{equation}\begin{gathered}
\label{7.3}y_{zzzzz}- 10\,y\,y_{zzz}- 20\,y_z\, y_{zz}
 + 30\,y^2\,y_z -z\,y_{{z}}-2\,y=0
\end{gathered}\end{equation}

Our results are summarized in the following theorem.
\begin{theorem}
\label{T:7.1.} Let $Y(z)$ be a solution of the equation

\begin{equation}
\label{7.7}Y_{{{ zzzzz}}}-40\,YY_{{{
zz}}}\,Y_{{z}}-10\,{Y}^{2}\,Y_{{{zzz}}}-10\,Y_{{z}}^{3}+30\,{Y}^{4}Y_{{z}}-Y-z\,Y_{{z}}=0
\end{equation}
Then

\begin{equation}
\label{7.8}y(z)=\,\pm\,Y_z-Y^2
\end{equation}
is a solution of equation \eqref{7.3}.

\end{theorem}

\begin{proof}The following points correspond to the monomials of this equation:
$M_1=(-5,1)$, $M_2=(-3,2)$, $M_3=(-3,2)$, $M_4=(-1,3)$, $M_5=(0,1)$
and $M_6=(0,1)$. These points generate the support of equation
\eqref{7.3}: $Q_1=M_1=(-5,1)$, $Q_2=M_4=(-1,3)$, $Q_3=M_5=M_6=(0,1)$
and $Q_4=M_2=M_3=(-3,2)$. Now we can plot the Newton polygon $L_1$
(see Fig. 5)

\begin{figure}[h]
 \centerline{\epsfig{file=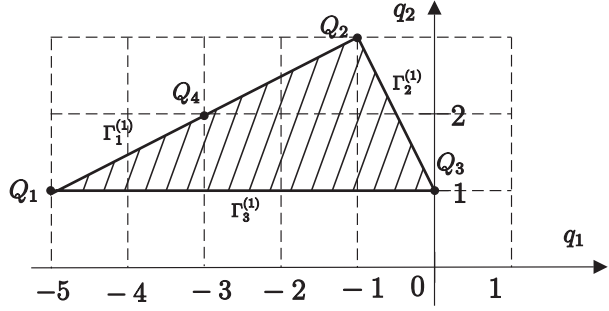,width=80mm}}
 \caption{Polygon corresponding to the differential equation of  \eqref{7.3}.}\label{fig5:z_post}
\end{figure}

Solutions of equation \eqref{7.3} have the second order pole. Thus
we can make the following transformation

\begin{equation}\begin{gathered}
\label{7.4}y=A_{{1}}Y+A_{{2}}{Y}^{2}+A_{{3}}Y_{{z}},
\end{gathered}\end{equation}
where $Y(z)$ is a function of the first order pole. Substituting
\eqref{7.4} into the equation studied, we obtain
\begin{equation}\begin{gathered}
\label{7.5} A_{{1}}Y_{{{\it zzzzz}}}+A_{{2}}{Y}^{2}_{{{\it
zzzzz}}}+A_{{3}}Y_{{{ \it zzzzzz}}}-\\
\\
-10\, \left( A_{{1}}Y+A_{{2}}{Y}^{2}+A_{{3}}Y_{{z}}
 \right)  \left( A_{{1}}Y_{{{\it zzz}}}+A_{{2}}{Y}^{2}_{{{zzz}}}+A
_{{3}}Y_{{{\it zzzz}}} \right) -\\
\\
-20\, \left( A_{{1}}Y_{{z}}+{A_2}\,{ Y}^{2}_{{z}}+A_{{3}}Y_{{{\it
zz}}} \right)  \left( A_{{1}}Y_{{{\it zz} }}+A_{{2}}{Y}^{2}_{{{\it
zz}}}+A_{{3}}Y_{{{\it zzz}}} \right) +\\
\\
+30\,
 \left( A_{{1}}Y+A_{{2}}{Y}^{2}+A_{{3}}Y_{{z}} \right) ^{2} \left( A_{
{1}}Y_{{z}}+A_{{2}}{Y}^{2}_{{z}}+A_{{3}}Y_{{{\it zz}}} \right) -\\
\\
-z
 \left( A_{{1}}Y_{{z}}+A_{{2}}{Y}^{2}_{{z}}+A_{{3}}Y_{{{\it zz}}}
 \right) -2\,A_{{1}}Y-2\,A_{{2}}{Y}^{2}-2\,A_{{3}}Y_{{z}}=0.
\end{gathered}\end{equation}

\begin{figure}[h]
 \centerline{\epsfig{file=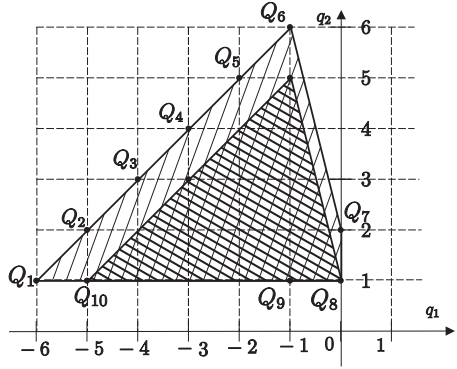,width=60mm}}
 \caption{Polygons of the differential equations \eqref{7.5} and \eqref{7.6}.}\label{fig6:z_post}
\end{figure}

The support of this equation is defined by ten points: $Q_1=(-6,1)$,
$Q_2=(-5,2)$, $Q_3=(-4,3)$, $Q_4=(-3,4)$, $Q_5=(-2,5)$,
$Q_6=(-1,6)$, $Q_7=(0,2)$, $Q_8=(0,1)$, $Q_9=(-1,1)$,
$Q_{10}=(-5,1)$. In this case, the polygon $L_2$ is the quadrangle
presented in Figure 6. Studying this polygon we can find the polygon
$L_3$ for the simplest equation. It is the triangle also presented
in Figure 6. Consequently the simplest equation for \eqref{7.5} is
\begin{equation}\begin{gathered}
\label{7.6}E_5[Y]\,=\,Y_{{{\it zzzzz}}}+m_1\,YY_{{{\it
zz}}}Y_{{z}}+m_2\,{Y}^{2}Y_{{{\it zzz}}
}+m_3\,{Y_{{z}}}^{3}+\\
\\
+m_4\,{Y}^{4}Y_{{z}}+m_5\,Y+m_6\,z\,Y_{{z}}=0,
\end{gathered}\end{equation}
where
$m_{1},\,\,\,\,m_{2},\,\,\,\,\,m_{3}\,\,\,\,\,m_{4},\,\,\,\,m_{5}$
and $m_{6}$ are unknown parameters to be found.

 Substituting
\begin{equation}\begin{gathered}
\label{7.9} Y_{zzzzz}\,=\,E_5[Y]-m_1\,YY_{{{\it
zz}}}Y_{{z}}-m_2\,{Y}^{2}Y_{{{\it zzz}}
}-m_3\,{Y_{{z}}}^{3}-\\
\\
-m_4\,{Y}^{4}Y_{{z}}-m_5\,Y-m_6\,z\,Y_{{z}}
\end{gathered}\end{equation}
into equation \eqref{7.5} and equating coefficients at different
powers of $Y(z)$ to zero yields algebraic equations for parameters
$m_1$, $m_2$, $m_3$, $m_4$, $m_5$ and $m_6$. As a result we get
\begin{equation}\begin{gathered}
\label{7.10}m_1=-40,\,\,\,\,m_2=-10,\,\,\,\,m_3=-10,\,\,\,\,m_4=30,\,\,\,\,\\
\\
m_5=-1,\,\,\,\,m_6=-1,
\end{gathered}\end{equation}
and the relation
\begin{equation}
\label{7.11}M_{5}[y(Y)]\,=\hat{R}\,E_{5}[Y],
\end{equation}
where $\hat{R}$ is a differential operator. This completes the
proof.
\end{proof}

Equation \eqref{7.7} can be integrated in $z$ and we get the fourth
-- order analogue to the second Painlev\'{e} equation

\begin{equation}\begin{gathered}
\label{7.12}Y_{{{zzzz}}}-10\,{Y}^{2}\,Y_{{{zz}}}-10\,Y\,{Y_{{z}}}^{2}+6\,{Y}^{
5}-z\,Y \,-\,\beta=0,
\end{gathered}\end{equation}
where $\beta$ is an arbitrary constant. Thus we have expressed
solutions of equation \eqref{7.3} (and consequently of equation
\eqref{7.1}) through solutions of \eqref{7.12}.

\section{Self similar solutions of the fifth -- order modified Korteveg -- de Vries equation}

Let us find self similar solutions of the fifth -- order modified
Korteveg -- de Vries equation
\begin{equation}
\label{8.1}u_t-10u^{2}u_{xxx}-40u_{x}u_{xx}-10u_{x}^{3}+30u^{4}u_{x}+u_{xxxxx}=0.
\end{equation}
It has the self similar solution \cite{Hone02, Kudryashov10,
Kudryashov11}
\begin{equation}
\label{8.2}u(x,t)={(5\,t)^{-1/5}}w(z),\,\,\,\,\,z={x}\,{(5\,t)^{-1/5}},
\end{equation}
where $w(z)$ satisfies
\begin{equation}
\label{8.3}M_4[w]=w_{zzzz}-10\,w^2w_{zz}-10\,ww_{z}^{2}+6\,w^5-z\,w-\beta=0.
\end{equation}

Let us present our result in the theorem.
\begin{theorem}
\label{T:8.1.} Let $Y(z)$ be a solution of the equation
\begin{equation}
\label{8.7}Y_{zzz}+2\,b\,Y\,Y_{zz}-b\,Y_{z}^{2}-6\,b^2\,Y^2\,Y_z-3\,b^3\,Y^4+\frac{z}{2\,b}=0.
\end{equation}
Then
\begin{equation}
\label{8.8}w(z)=\,\pm\,b\,Y(z)
\end{equation}
is a solution of the equation \eqref{8.3}, provided that
$\beta=\mp1/2$.

\end{theorem}

\begin{proof}To begin with the Newton polygon corresponding to the equation \eqref{8.3}
should be found. The following points: $M_1=(-4,1)$, $M_2=(-2,3)$,
$M_3=(-2,3)$, $M_4=(0,5)$, $M_5=(1,1)$, $M_6=(0,0)$ are assigned to
the monomials of the studied equation. The support of the equation
contains five points $Q_1=M_1,\,\,\, Q_2=M_4,\,\,\, Q_3=M_5 ,\,\,\,
Q_4=M_6$ è $Q_5=M_2=M_3$. Their convex hull is the quadrangle $L_1$
with four vertexes $\Gamma_j^{(0)}=Q_j\,(j=1,2,3,4)$ and four edges
$\Gamma_1^{(1)}=[Q_1,Q_2],\,\Gamma_2^{(1)}=[Q_2,Q_3],\,\Gamma_3^{(1)}=[Q_3,Q_4],\,\Gamma_4^{(1)}=[Q_1,Q_4]$
(fig. 7).

\begin{figure}[h]
 \centerline{\epsfig{file=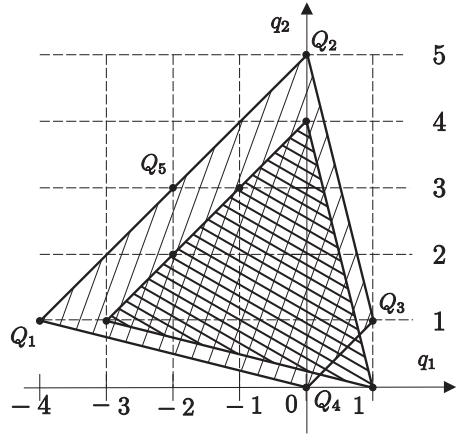,width=60mm}}
 \caption{Polygons of the differential equations \eqref{8.3} and \eqref{8.5}.}\label{fig7:z_post}
\end{figure}
Leadings members of the equation \eqref{8.1} are located on the edge
$\Gamma_1^{(1)}=[Q_1,Q_2]$. Substituting $w(z)=a_0\,z^p$ into the
equation
\begin{equation}
\label{8.3a}w_{zzzz}-10\,w^2w_{zz}-10\,ww_{z}^{2}+6\,w^5=0,
\end{equation}
we get fours families of power asymptotics for solutions of
\eqref{8.3}: $(p,\,a_0)=(-1,\,1)$, $(-1,\,-2)$, $(-1,\,-3)$, and
$(-1,\,4)$. Again taking into account the pole order of solutions
and the results of section 2, we can look for exact solutions of the
equation \eqref{8.3} in the form
\begin{equation}
\label{8.4}w(z)=A_{1}\,Y(z),\,\,\,A_1\neq 0,
\end{equation}
where $Y(z)$ is a solution of the following simplest equation
\begin{equation}
\label{8.5}Y_{zzz}-a \,YY_{zz}-b \,Y_{z}^{2}-c\, Y^{2}Y_z-d\,
Y^4-m\, z=0.
\end{equation}
It is important to mention that the transformation \eqref{8.3} does
not change the quadrangle $L_1$ and the Newton polygon $L_3$ for the
simplest equation \eqref{8.4} possesses all necessary properties. In
other words $L_1\equiv L_2$ and $L_3$ is the triangle (see Fig. 7).
The parameters $a$,\, $b$, \,$c$, \,$d$, and $m$ can be found.

 Substituting \eqref{8.4} and
\begin{equation}\begin{gathered}
\label{8.9}Y_{zzz}\,=\,E_3[Y]\,+a \,YY_{zz}+b \,Y_{z}^{2}+c\,
Y^{2}Y_z+d\, Y^4+m\, z=0
\end{gathered}\end{equation}
into equation \eqref{8.3} and equating coefficients at different
powers of $Y(z)$ to zero yields algebraic equations for parameters
$a$, $b$, $c$, $d$, $m$, $\beta$ and $A_1$. Finally, we obtain

\begin{equation}\begin{gathered}
\label{8.10}a=-2\,b,\,\,\,c=6\,b^2,\,\,\,d=3\,b^3,\,\,\,m=-\frac{1}{2\,b},\\
\\
A_{1,2}=\pm\,b,\,\,\,\,\,\beta=\mp\,\frac12
\end{gathered}\end{equation}
and the relation \cite{Kudryashov10}
\begin{equation}
\label{8.11}M_{4}[w(Y)]\,=\hat{R}\,E_{3}[Y],
\end{equation}
where $\hat{R}$ is a differential operator. This completes the
proof.
\end{proof}

Consequently there exist solutions of the studied equation expressed
through the function $Y(z)$ at $\beta =1/2$ or $\beta=-1/2$. Let us
show that solutions of the equation \eqref{8.7} are associated with
solutions of the first Painlev\'{e} equation
\begin{equation}
\label{8.12}v_{zz}=C\,v^2+D\,z.
\end{equation}
Making the transformation
\begin{equation}
\label{8.13}v(z)\,=\,A\,w^2\,+\,B\,w_z
\end{equation}
in the equation \eqref{8.8}, yields the equation
\begin{equation}
\label{8.14} B\,w_{zzz}+2\,A\,ww_{zz}+(2A-CB^2)
\,w_{z}^{2}-2\,C\,A\,B w^{2}w_z-C\,A^2\, w^4-D\, z=0.
\end{equation}
We see that the Newton polygon corresponding to this equation
coincides with the triangle $L_3$ at Figure 6. Combining \eqref{8.7}
and \eqref{8.10}, we obtain
\begin{equation}
\label{8.15} w(z)\equiv Y(z),
\end{equation}
provided that $A=b$, $B=1$, $C=3b$, and $D=-1/2b$. In other words,
if $Y(z)$ is a solution of \eqref{8.7}, then $v(z)=b\,Y^2+Y_z$ is a
solution of
\begin{equation}
\label{8.16} v_{zz}=3\,b\,v^2-\frac1{2b}\,z.
\end{equation}

We would like to mention that without loss of generality it can be
set $b=\pm1$.

\section{Solitary waves of the sixth order nonlinear evolution equation}

Let us find exact solutions of the sixth order nonlinear evolution
equation that is applied at description of turbulent processes
\cite{Kudryashov12, Nikolaevsky01}

\begin{equation}
\label{9.1}u_t + uu_x +\beta \,u_{xx} +\delta\, u_{xxxx}.
+\varepsilon \,u_{xxxxxx}=0.
\end{equation}

It is known that Kuramoto -- Sivashinskiy equation and the Ginzburg
-- Landau equation that are used at description of turbulence are
non-integrable ones because they do not pass the Painlev\'e test
\cite{Kudryashov01}. However, these equations have a list of special
solutions \cite{Kudryashov01, Kudryashov02, Kudryashov05,
Eremenko01, Hone02}.

Equation \eqref{9.1} does not pass the Painlev\'e test as well and
thus this equation is also non-integrable. However one can expect
that equation \eqref{9.1} has some special solutions. This equation
admits the travelling waves reduction
\begin{equation}
\label{9.2}u(x,t)=y(z), \qquad z=x-C_0 t,
\end{equation}
where $y(z)$ satisfies the equation
\begin{equation}
\label{9.5}C_1 -C_0 y +\frac12 y^2 +\beta y_z +\delta y_{zzz} +
\varepsilon y_{zzzzz}=0.
\end{equation}
$C_1$ is a constant of integration. As a result we get the following
theorem.
\begin{theorem}
\label{T:9.1.} Let $Y(z)$ be a solution of the equation

\begin{equation}
\label{9.6}Y_{z}\,+
\,Y^{2}-\alpha_k\,=\,0,\,\,\,\,\,\,(k=1,2,...,6).
\end{equation}
Then
\begin{equation}
\begin{gathered}
\label{9.7}y \left( z \right) =30240\,\varepsilon\, Y ^{5}+ \left(
{\frac {2520}{11}}\,\delta-50400\,\varepsilon\,
\alpha_k \right)  Y^{3}+\\
\\+ \left( -{ \frac
{2520}{11}}\,\delta\,\alpha_k+20160\,\varepsilon\,{\alpha_k}^{2}+{
\frac {1260}{251}}\,\beta-{\frac {12600}{30371}}\,{\frac
{{\delta}^{2} }{\varepsilon}}\right) Y +C_0
\end{gathered}\end{equation}
is a solution of the equation \eqref{9.5} provided that
\begin{equation}\begin{gathered}
\label{9.8} C_1={\frac {4112640}{11}}\,{\frac
{{\delta}^{5}{w_k}^{4}}{{\varepsilon }^{3}}}-9999360\,{\frac
{{\delta}^{5}{w_k}^{5}}{{\varepsilon}^{3}}} -{\frac
{5080320}{251}}\,{\frac
{{\delta}^{3}{w_k}^{3}\beta}{{\varepsilon}^{2}}}-\\
\\
-{ \frac {55460160}{30371}}\,{\frac
{{\delta}^{5}{w_k}^{3}}{{\varepsilon}^{3}} }+{\frac
{660240}{2761}}\,{\frac {{\delta}^{3}{w_k}^
{2}\beta}{{\varepsilon}^{2}}}-{\frac {25200}{30371}}\,{\frac
{{\delta}^{5 }{w_k}^{2}}{{\varepsilon}^{3}}}-\\
\\
-{\frac {1260}{251}}\,{\frac {{\beta}^{2}
\delta\,w_k}{\varepsilon}}+{\frac {12600}{30371}}\,{\frac
{\beta\,{\delta}^ {3}w_k}{{\varepsilon}^{2}}}+\frac12\,{C_0}^{2},
\end{gathered}\end{equation}
\begin{equation}
\label{9.9}\beta=-{\frac {213811840\,{\varepsilon}^{3}{\alpha_k}^{3}
-10204656\,\delta\,{\varepsilon}^{2}{\alpha_k}^{2}-2045\,{\delta}^{3}-92400
\,{\delta}^{2}\varepsilon\,\alpha_k}{121 \varepsilon\, \left(
9240\,\varepsilon\, \alpha_k+79\,\delta \right) }},
\end{equation}
where $\alpha_k$ and $w_k$ are found from

\begin{equation}
\label{9.12}\alpha_k=\frac{\delta\,
w_k}\varepsilon\,\,\,\,\,\,(k=1,2,...,6),
\end{equation}
\begin{equation}
\begin{gathered}
\label{9.13} w_1=-{\frac {1}{220}}, \quad w_2=-{\frac {5}{176}},
\quad w_3=-{\frac {1}{440}},
\end{gathered}\end{equation}

\begin{equation}
\begin{gathered}
\label{9.14}w_4 =\frac1{52800} \left(557 - \frac{46031}{m} +
m\right),
\end{gathered}\end{equation}

\begin{equation}
\begin{gathered}
\label{9.15}w_{5,6} =\frac1{52800} \left(\frac{46031}{2m} - \frac m2
+ 557 \pm \frac{i\sqrt{3}}2 \left( m+ \frac{46031}{m}\right)
\right),
\end{gathered}
\end{equation}

\begin{equation}
\begin{gathered}
\label{9.16}m=(113816753 +1260 \sqrt{8221079733})^\frac13 \approx
610,966.
\end{gathered}\end{equation}

\end{theorem}

\begin{proof}

The following points correspond to the monomials of equation
\eqref{9.5}: $Q_1=(-5,1)$, $Q_2=(0,2)$, $Q_3=(0,0)$, $Q_4=(-3,1)$,
and $Q_5=(-1,1)$. The Newton polygon $L_1$  of the equation is the
triangle (Fig. 8).

\begin{figure}[h]
 \centerline{\epsfig{file=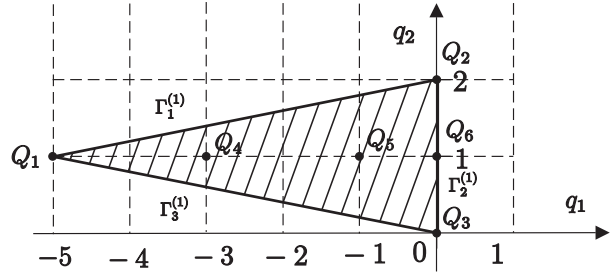,width=80mm}}
 \caption{Polygon corresponding to the differential equation \eqref{9.5}.}\label{fig8:z_post}
\end{figure}

Solutions of equation \eqref{9.5} have the fifth -- order pole. Let
us express them through a function $Y(z)$ possessing the first --
order pole
\begin{equation}
\label{9.18}y (z) = A_0+ A_1 Y  +A_2
  Y  ^{2}+ A_3  Y
 ^{3}+A_4 Y  ^ {4}+A_5  Y  ^{5}.
\end{equation}
Substituting transformation \eqref{9.7} into \eqref{9.5}, we get new
equation and the corresponding polygon $L_2$ in the form of
quadrangle (Fig.9). One of the suitable polygons $L_3$ is the
triangle in Fig. 9. This triangle is assigned to the Riccati
equation with constant coefficients
\begin{equation}
\label{9.19}E_1[Y]=Y_z +Y^2 -\alpha =0.
\end{equation}

\begin{figure}[h]
 \centerline{\epsfig{file=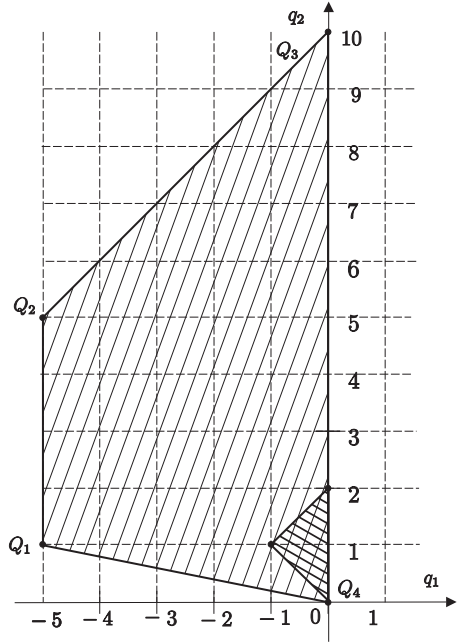,width=60mm}}
 \caption{Polygon corresponding to the transformed differential equation \eqref{9.5} and to the simplest differential equation \eqref{9.19}.}\label{fig9:z_post}
\end{figure}
 Substituting
\begin{equation}
\label{9.20} Y_z\,=\,E_1[Y]\,-Y^2+\alpha_k
\end{equation}
into equation \eqref{9.6} and equating coefficients at different
powers of $Y(z)$ to zero yields algebraic equations for parameters
$A_5$, $A_4$, $A_3$, $A_2$, $A_1$, $A_0$, $C_1$, $\beta$, $\alpha_k$
and $w_k$. Solving these equations we have

\begin{equation}
\begin{gathered}
\label{9.21}A_5=30240\,\varepsilon, \quad A_4=0, \quad A_3=\frac
{2520\delta}{11}-50400\varepsilon \alpha, \quad A_2=0,\\ \\
 A_1=-{\frac
{2520}{11}}\,\delta\,\alpha+20160\,\varepsilon\,{\alpha} ^{2}+{\frac
{1260}{251}}\,\beta-{\frac {12600}{30371}}\,{\frac {{
\delta}^{2}}{\varepsilon}}, \quad A_0=C_0.
\end{gathered}
\end{equation}
For the parameters $C_1$, $\beta$, $\alpha_k$, and $w_k$ we obtain
expressions \eqref{9.8}, \eqref{9.9}, \eqref{9.12}, \eqref{9.13},
\eqref{9.14}, and \eqref{9.16}. We also see that the following
relation
\begin{equation}
\label{9.22}M_5[y(Y)]\,=\hat{R}\,E_1[Y],
\end{equation}
holds, where $\hat{R}$ is a differential operator. This completes
the proof.
\end{proof}

The general solution of equation \eqref{9.6} is the following
\begin{equation} \label{9.23}Y(z)=\sqrt{\alpha_k}\tanh
\left(\sqrt{\alpha_k} z +\varphi_0\right),\,\,\,\,\,\,(k=1,2,...,6).
\end{equation}

Substituting \eqref{9.23} into \eqref{9.7} and taking into account
that $ \alpha_k=w_k \delta/\varepsilon$ $(k=1, ..., 6)$ we get six
different solitary waves of equation \eqref{9.5}.

\section{Exact periodic solutions of equation \eqref{9.1}}

It was mentioned in the previous section that solutions of equation
\eqref{9.5} have the fifth order pole. Thus let us make the
following transformation
\begin{equation}
\label{10.1}y \left( z \right) =B_1+ B_2 R ( z ) + B_3R_z + B_4
R^{2}+B_5RR_z,
\end{equation}
where $B_k \,\,(k=1, ..., 5)$ are constants and $R=R(z)$ is a
function of the second order pole. Now let us prove the following
theorem.

\begin{theorem}
\label{T:10.1.} Let $R(z)$ be a solution of the equation

\begin{equation}\begin{gathered}
\label{10.4} R^2_z=-2\, R ^{3}+a  R ^ {2}-\frac16{a}^{2}R +{\frac
{1}{726}}{\frac {R  {\delta}^{2}}{{\varepsilon}^{2}}}\pm
 {\frac {1}{2541}}{\frac {R  \sqrt {21}{\delta}^{2}}{{\varepsilon}^{2}}}+\\ \\{ \frac
{1}{108}}\,{a}^{3}+{\frac {13}{359370}}\,{\frac {{\delta}^{3}}{{
\varepsilon}^{3}}}\pm{\frac {1}{119790}}\,{\frac {\sqrt
{21}{\delta}^{3}}{{ \varepsilon}^{3}}}-{\frac {1}{4356}}\,{\frac
{a{\delta}^{2}}{{\varepsilon}^{ 2}}}\mp{\frac {1}{15246}}\,{\frac
{a\sqrt {21}{\delta}^{2}}{{\varepsilon}^{ 2}}}.
\end{gathered}\end{equation}
Then

\begin{equation}
\label{10.5}y(z)=C_0 + 630 \left(\varepsilon a +\frac\delta {11}
-6\varepsilon R\right) R_z
\end{equation}
is a solution of equation \eqref{9.1},provided that

\begin{equation}
\label{10.5_1}\beta={\frac {10}{121}}\,{\frac
{{\delta}^{2}}{\varepsilon}},
\end{equation}

\begin{equation}
\begin{gathered}
\label{10.6}C^{(1,2)}_1=-{\frac {10854}{161051}}\,{\frac
{{\delta}^{5}}{{\varepsilon}^{ 3}}}+\frac12{C_0}^{2}\mp{\frac
{2484}{161051}}\,{\frac {\sqrt {21}{
\delta}^{5}}{{\varepsilon}^{3}}}.
\end{gathered}
\end{equation}

\end{theorem}

\begin{proof}
Substituting \eqref{10.1} into equation \eqref{9.5} we get a new
equation of the sixth order. The Newton polygon $L_2$ of this
equation is presented in Figure 10. Again we should construct
suitable polygon $L_3$. Let us take the triangle corresponding to
the equation for the elliptic function

\begin{figure}[h]
 \centerline{\epsfig{file=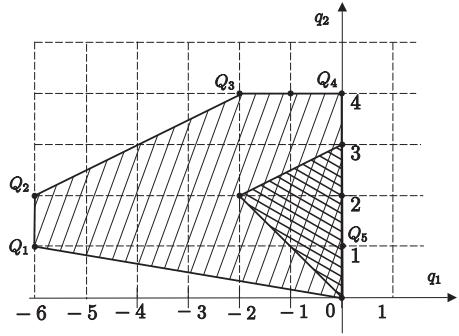,width=60mm}}
 \caption{Polygons corresponding to the transformed differential equation \eqref{9.5} and to the simplest differential equation \eqref{10.2}.}\label{fig10:z_post}
\end{figure}

\begin{equation}
\label{10.2}E_1[R]\,=\,R^2_z +2R^3 -aR^2 -2bR -d=0.
\end{equation}

This triangle satisfies necessary requirements. Substituting
\begin{equation}
\label{10.7}R_z^2=E_1[R]\,-\,2\,R^3\,+\,a\,R^2 \,+\,2\,b\,R\,+\,d,
\end{equation}

\begin{equation}
\label{10.8}R_{zz}=\frac{1}{2\,R_z}\,\frac{\partial
E_{1,z}[R]}{\partial z}\,-\,3\,R^2\,+\,a\,R\,\,+\,b
\end{equation}
into equation \eqref{9.5} and equating coefficients at different
powers of $R(z)$ and $R_z$ to zero yields algebraic equations for
the coefficients $B_5$, $B_4$, $B_3$, $B_2$, $B_1$, and $B_0$. As a
result we obtain the following expressions

\begin{equation}\begin{gathered}
\label{10.9}B_5\,=\,-3780\,\varepsilon,\,\quad \,B_4=0,\quad \,
B_3\,=630\,\varepsilon\,a+{\frac
{630}{11}}\,\delta, \\
\\
\,B_2=0,\,\quad \,B_1= C_0,\end{gathered}\end{equation}

\begin{equation}
\label{10.10}\beta={\frac {10}{121}}\,{\frac
{{\delta}^{2}}{\varepsilon}},
\end{equation}

\begin{equation}
\begin{gathered}
\label{10.11} b_{1,2}=-\frac1{12} {a}^{2}+{\frac {1}{1452}}\,{\frac
{{\delta}^{2}}{{ \varepsilon}^{2}}}\pm {\frac {1}{5082}}\,{\frac
{\sqrt {21}{\delta}^{2}}{{ \varepsilon}^{2}}},
\end{gathered}
\end{equation}

\begin{equation}
\begin{gathered}
\label{10.12}d_{1,2}={\frac {1}{108}}\,{a}^{3}+{\frac
{13}{359370}}\,{\frac {{ \delta}^{3}}{{\varepsilon}^{3}}}\pm{\frac
{1}{119790}}\,{\frac {\sqrt {21}{
\delta}^{3}}{{\varepsilon}^{3}}}-\\
\\
-{\frac {1}{4356}}\,{\frac {a{\delta}^{2}
}{{\varepsilon}^{2}}}\mp{\frac {1}{15246}}\,{\frac {a\sqrt
{21}{\delta}^{2} }{{\varepsilon}^{2}}},
\end{gathered}
\end{equation}
and the relation

\begin{equation}
\label{10.13}M_{6}[y(R)]\,=\hat{R}\,E_{1}[R],
\end{equation}
where $\hat{R}$ is a differential operator. This completes the
proof.
\end{proof}

If $R_1,R_2$, and $R_3$ such that $R_1\geq R_2 \geq R_3$ are real
roots of the equations

\begin{equation}
\begin{gathered}
\label{10.14} 2R^3 -aR^2 +\left(\frac16 a^2 -
\frac{\delta^2}{726\varepsilon^2} \mp
\frac{\delta^2 \sqrt{21}}{2541\varepsilon^2}\right)R-\frac1{108} a^3-\\ \\
-\frac{13}{359379} \frac{\delta^3}{\varepsilon^3} \mp
\frac{\delta^3\sqrt{21}}{119790\varepsilon^3}
+\frac{a\delta^2}{4356\varepsilon^2} \pm\frac
{a\delta^2\sqrt{21}}{15246\varepsilon^2}=0,
\end{gathered}
\end{equation}
then the general solution of \eqref{10.4} can be written as
\begin{equation}
\begin{gathered}
\label{10.15} R(z) =R_2 + (R_1-R_2)\,
\textup{cn}^2(z\sqrt{\frac{R_1-R_3}{2}},S),\quad
S^2=\frac{R_1-R_2}{R_1-R_3}.
\end{gathered}
\end{equation}

 Thus we have found several solutions of equation \eqref{9.1} at different values of
the parameters. These solution are solitary and periodic waves (see
expressions \eqref{9.7} and \eqref{10.5}).

\section{Conclusion}

In this paper a new method for finding exact solutions of nonlinear
differential equations is presented. Our aim was to express
solutions of the equation studied through solutions of the simplest
equations. To search the simplest equation we applied the power
geometry and constructed the Newton polygons of nonlinear
differential equations. The method is more powerful than many other
methods because the structure of a solution is not fixed but can be
found using the stated algorithm. Thus one can obtain quite general
classes of exact solutions. It is important to mention that
transformations between nonlinear differential equations can be also
found with a help of our approach. As an example of our method
application exact solutions of some nonlinear differential equations
were found. In particular, solutions of the fourth -- order
evolution equation \eqref{4.1} were expressed in terms of the Airy
functions; the connection between self -- similar solutions of the
fifth -- order Korteveg -- de Vries equation \eqref{7.1} and the
fifth -- order modified Korteveg -- de Vries equation \eqref{8.1}
were constructed; one -- parametric family of exact solutions for
the generalized Kuramoto - Sivashinsky equation \eqref{e:5.0} was
found. Besides that we obtained solitary waves and periodic
solutions of the sixth -- order evolution equation \eqref{9.1} that
is used at description of turbulence \cite{Nikolaevsky01, Cox01}.

\section {\textbf{Acknowledgments}}

This work was supported by the International Science and Technology
Center under Project No B 1213.

\end{document}